\documentclass[twocolumn,showpacs,preprintnumbers]{revtex4}
\usepackage{amssymb}
\usepackage{amsmath}
\usepackage{graphicx}
\usepackage{dcolumn}
\usepackage{bm}
\usepackage{color}

\begin{document}

\title{Reversible ratchet effects in a narrow superconducting ring}
\author{Ji Jiang$^{1,2}$, Yong-Lei Wang$^{3}$, M. V. Milo\u{s}evi\'{c}$^{4}$,
Zhi-Li Xiao$^{5,6}$, F. M. Peeters$^{4}$, and Qing-Hu Chen$^{1,2}$}

\address{
$^{1}$Zhejiang Province Key Laboratory of Quantum Technology and Device, Department of Physics, Zhejiang University, Hangzhou 310027, China \\
$^{2}$Collaborative Innovation Center of Advanced Microstructures, Nanjing University, Nanjing 210093, China \\
$^{3}$Research Institute of Superconductor Electronics, School of Electronic Science and Engineering, Nanjing University, Nanjing 210093, China \\
$^{4}$Department of Physics, University of Antwerp, 2020 Antwerpen, Belgium\\
$^{5}$Materials Science Division, Argonne National Laboratory, Argonne, Illinois 60439, USA \\
$^{6}$Department of Physics, Northern Illinois University, DeKalb, Illinois 60115, USA
}\date{\today}

\begin{abstract}
We study the ratchet effect in a narrow pinning-free superconductive ring
based on time-dependent Ginzburg-Landau (TDGL) equations. Voltage responses
to external dc an ac currents at various magnetic fields are studied. Due to asymmetric
barriers for flux penetration and flux exit in the ring-shaped superconductor,
the critical current above which the flux-flow state is
reached, as well as the critical current for the transition to the normal state, are
different for the two directions of applied current. These effects cooperatively
cause ratchet signal reversal at high magnetic fields, which has not been reported
to date in a pinning-free system. The ratchet signal found here is
larger than those induced by asymmetric pinning potentials. Our results also
demonstrate the feasibility of using mesoscopic superconductors to employ superconducting
diode effect in versatile superconducting devices.
\end{abstract}
\pacs{74.25.Wx, 74.25.Uv,74.78.Na,  74.40.Gh}
\maketitle

\section{Introduction}

In type-II superconductors, magnetic flux penetrates into the sample
under the magnetic field above the lower critical field, forming quantized
magnetic flux lines known as Abrikosov vortices. When applying a sufficiently
large current, vortices are driven across the sample due to Lorentz
force, resulting in finite voltage signals. If the vortex dynamics
differs with respect to the polarity of applied current, the associated voltage would be different,
producing the vortex ratchet effect. Vortex ratchet systems not only provide a convenient
platform to investigate the fundamental vortex dynamics, but also are applicable
in superconductive circuits. In that respect, manipulating
vortices using ratchet systems has been demonstrated in many experiments \cite%
{lee1999reducing,wambaugh1999superconducting,olson2001collective,villegas2003superconducting,vlasko2017magnetic,zhu2004controlling,veshchunov2016optical}%
. Ratchet effect can be employed to remove undesirable vortices
trapped in superconductors, to thereby improve the performance of superconducting
devices. Recently, the nonreciprocal charge transport has been observed in
various superconductors with noncentrosymmetric or chiral structures \cite%
{qin2017superconductivity,yasuda2019nonreciprocal,wakatsuki2017nonreciprocal}%
. More recently, the superconducting diode effect that has zero resistance
for only one direction of the current has been realized in a noncentrosymmetric superlattice
by stacking three kinds of superconducting elements \cite%
{ando2020observation}, which may pave the way for potential applications in
low dissipative electronic circuits.

Vortex ratchet systems are typically realized by introducing asymmetric pinning
potentials in the superconducting samples to fine tune the vortex dynamics
\cite%
{rouco2015geometrically,zhu2003controllable,adami2015onset,reichhardt2015reversible,reichhardt2016transverse,reichhardt2017pinning}%
. Pinning-free superconductors of special geometries could also serve as the
vortex ratchet systems. In a superconducting sample with
asymmetric edges, the vortex dynamics can be affected by
intrinsic edge barriers. Recently, such kinds of pinning-free vortex
ratchet superconducting systems have also attracted much attention \cite%
{ji2016vortex,ji2017ratchet,berdiyorov2012spatially,hortensius2012critical,schildermans2007voltage,adami2013current,vodolazov2005superconducting,berdiyorov2010vortices}%
. Ratchet systems without artificial pinning centers distinguish
themselves with the ability to produce a stable and strong rectifying effect
\cite{ji2016vortex,ji2017ratchet}. In practice, superconductors with specifically
targeted asymmetric geometry have been widely used in experiments   as single-photon
detectors, parametric amplifiers and superconducting quantum interference
devices(SQUIDs) \cite%
{halbertal2016nanoscale,day2003broadband,hortensius2012critical,mittendorff2013ultrafast}%
. On the other hand, superconducting nanowires and nanoribbons are the key
components in these advanced superconducting circuits or devices.
Low-dimensional superconducting structures provide unique properties and
have been widely studied \cite%
{berdiyorov2012magnetoresistance,berdiyorov2012large,blatt1963shape,PhysRevLett.61.2137,PhysRev.164.498,PhysRevB.1.1054,PhysRevLett.95.116805,kenawy2020electronically}%
. To improve the performance and reliability of superconducting devices, it
is crucial to understand the rich vortex dynamics in low-dimensional
superconducting systems with specified geometries.

Although ratchet effects have been reported in many systems in the
literature, ratchet signal reversal is seldomly observed. Systems
with specially designed edges to date have been reported to give
strong and stable ratchet signals without sign reversal \cite%
{ji2016vortex,ji2017ratchet,berdiyorov2012spatially,hortensius2012critical,schildermans2007voltage,adami2013current,vodolazov2005superconducting}%
. On the other hand, the ratchet reversal in bulk or two-dimensional (2D)
superconducting samples with asymmetric pinning arrays has been reported in
both experiments and numerical simulations \cite%
{reichhardt2015reversible,reichhardt2016transverse,adami2015onset,wu2015special},
though these reversible ratchet signals strongly depended on the properties of the
applied pinning potentials (such as density and strength of the pinning sites).

In this paper, by means of numerical simulations we reveal strong and stable
reversible ratchet signals in a broad range of magnetic fields and external
currents, in a pinning-free, narrow superconducting ring. We
even find a superconducting-diode-like state, with vanishing resistance for only
one current direction. The paper is organized as follows. In Sec. II, we introduce
the time-dependent Ginzburg-Landau (TDGL) equations, which are
used to simulate the condensate dynamics in the presence of external dc and ac
currents at various magnetic fields. The main simulation results for ratchet
effects generated under dc and ac currents are presented and
analyzed in Sec. III and IV, respectively. The associated mechanism for the reversible
ratchet signals is also described in detail. Additional videos for detailed visualization of
the vortex motion are provided in the Supplemental Material \cite{Supplemental} to help understand
the vortex dynamics. Finally, conclusions are drawn in Sec. V.

\section{Numerical approach}

\begin{figure}[tbp]
\centerline{
\includegraphics[scale=0.23]{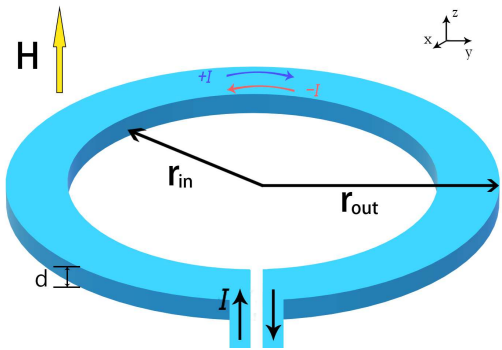}}
\caption{Schematic view of a narrow, pinning-free superconducting ring
with the outer radii $r_{out}$ and the inner radii $r_{in}$.
The thickness of the sample $d$ is assumed to be much smaller than the
penetration length $d << \protect\lambda$. A perpendicular magnetic field $H$
is applied. A clockwise (counter-clockwise) current is denoted by $+I$ ($-I$).}
\label{fig_sys}
\end{figure}

We use the TDGL theory to simulate a pinning-free superconducting ring.
An oblique view of the system is
shown in Fig. \ref{fig_sys}. The two TDGL equations for the
superconductive sample without any defects are given by \cite%
{bennemann2008superconductivity}:
\begin{equation}
\frac{\partial \psi }{\partial t}=(\nabla -i\mathbf{A})^{2}\psi +(1 - T -|\psi
|^{2}) \psi +\chi (\mathbf{r},t),  \label{tdgl_zp_1}
\end{equation}%
\begin{equation}
\sigma \frac{\partial \mathbf{A}}{\partial t}=Im(\psi ^{\ast }(\nabla -i%
\mathbf{A})\psi )-\kappa ^{2}\nabla \times \nabla \times \mathbf{A},
\label{tdgl_zp_2}
\end{equation}%
where $\psi $ is the superconducting order parameter, $\mathbf{A}$ is the vector
potential describing the magnetic field $\mathbf{B}=\nabla \times \mathbf{A}$, $\sigma $
is the conductivity in the normal state, and $\kappa =\lambda /\xi $ is the
Ginzburg-Landau (GL) parameter (here  $\lambda $ is the penetration depth and $%
\xi $ is the coherence length). $\chi (\mathbf{r},t)$ is introduced to mimic the
quantum fluctuations in the system \cite%
{PhysRevB.47.8016,berdiyorov2012large}. The length is scaled to coherence
length of zero temperature $\xi(0) $ and time to $t_{GL}=4\pi \lambda(0)
^{2}\sigma /c^{2}$, which is known as the GL relaxation time. The vector
potential is scaled so that the magnetic field is in units of the bulk upper
critical field $H_{c2}=\Phi _{0}/2\pi \xi(0) ^{2}$ where $\Phi _{0}$ is the
flux quantum. Equation (\ref{tdgl_zp_2}) is actually the
Maxwell equation governing the magnetic field and total current. The current
is in units of $I_{0}=\sigma \hbar w/2e\xi(0) t_{GL}$ where $w$ is the
width of the ring. To solve Eqs. (\ref{tdgl_zp_1} and  \ref{tdgl_zp_2}), we employ
the zero-potential scheme \cite%
{bennemann2008superconductivity,winiecki2002fast,berdiyorov2012magnetoresistance}%
, \textit{i.e.} the zero electric potential gauge: $\phi =0$.

In this work, we focus on the ring of width comparable to the vortex size, so
more than one row of vortices aligned along the ring are not energetically
favorable \cite{cordoba2019long}. We set $\kappa = 10$ in all of our simulations,
which is close to the value reported for typical MoGe superconducting samples \cite%
{wang2017parallel}. Referring to previous simulation works on wide ring-shaped
samples \cite{ji2017ratchet}, the outer and inner radii are set at $%
r_{out}=60\xi(0) $ and $r_{in}=48\xi(0) $, respectively, so the width of the
circular strip is $w=12\xi(0) $, slightly larger than the penetration depth.
The magnetic field is applied perpendicular to the sample as shown in
Fig. \ref{fig_sys}. When solving Eq. \ref{tdgl_zp_1}, the Neumann boundary condition
is applied at all sample edges. Considering that our superconducting sample is
extremely thin, we neglect the demagnetization effects and
apply the external current through the following boundary condition for the
vector potential $\mathbf{A}$ at the inner boundary: $%
\nabla \times \mathbf{A} = \mathbf{H} \mp \mathbf{H}_{I}$ for $\pm I$ respectively, where $%
H_{I} = 2 \pi I / c$ is the magnetic field induced by the applied current $I$.
The sign of $H_I$ defines the direction of the current flow in the system.
For notational simplicity, we denote clockwise (counter-clockwise) current
as $+I$ ($-I$).

We also consider the effect of Joule heating in the simulations, i.e.,
dissipation generated by moving vortices, where the system would have a
non-uniform temperature distribution. The dimensionless heat transfer
equation is used to describe the dynamics of thermal diffusion \cite%
{berdiyorov2012large,PhysRevB.71.184502}:
\begin{equation}
\nu \frac{\partial T}{\partial t}=\zeta \nabla ^{2}T+(\sigma \frac{\partial
\mathbf{A}}{\partial t})^{2}-\eta (T-T_{0}),  \label{heateq}
\end{equation}%
where $\nu $, $\zeta $,  $\eta $ are the heat capacity, heat conductivity of
the sample, and heat coupling to the environment, respectively, which we set
in simulations to $\nu =0.03$, $\zeta =0.06$, and $\eta =2\times 10^{-4}$.
The value of $\eta$ corresponds to intermediate heat removal \cite%
{PhysRevB.71.184502}. The values of $\nu$ and $\zeta$ are roughly
estimated, as employed in previous simulations of mesoscopic superconductors \cite%
{berdiyorov2012large,PhysRevB.71.184502}, where reliable results have been reported. The temperature $T$ in
the simulations is scaled by the superconducting critical
temperature $T_{c}$. $T_{0}=0.9T_{c}$ is the temperature of the environment (holder)
and is constant in all our simulations. Equations (\ref{tdgl_zp_1}), (\ref{tdgl_zp_2}),
and (\ref{heateq}) are solved self-consistently using Crank-Nicholson method \cite{winiecki2002fast}.

\section{Ratchet effects generated by dc currents}

\subsection{Phase diagram of the ratchet signal in terms of the current and
the field}

\begin{figure}[tbp]
\centering
\includegraphics[scale=0.5]{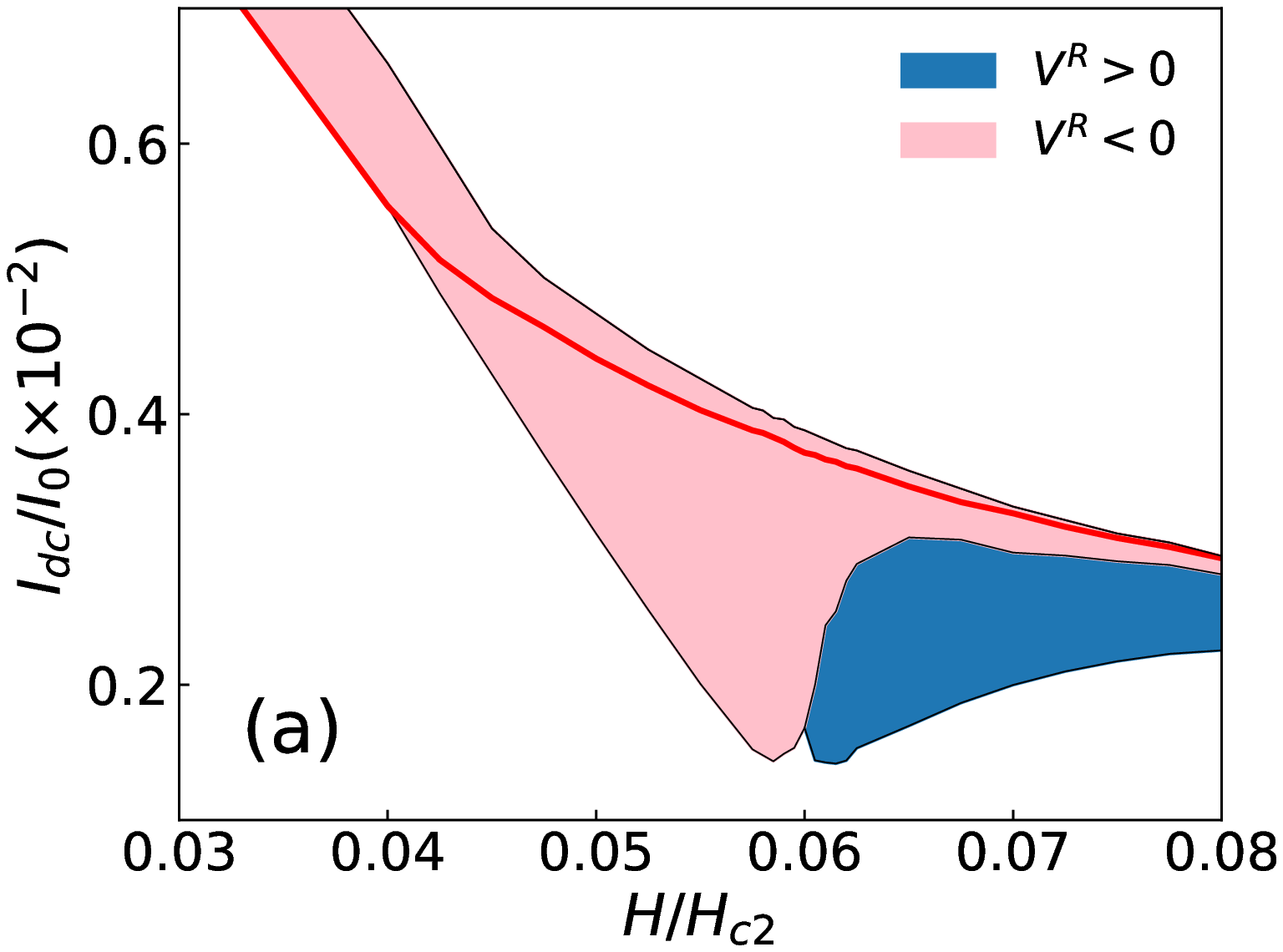} %
\includegraphics[scale=0.33]{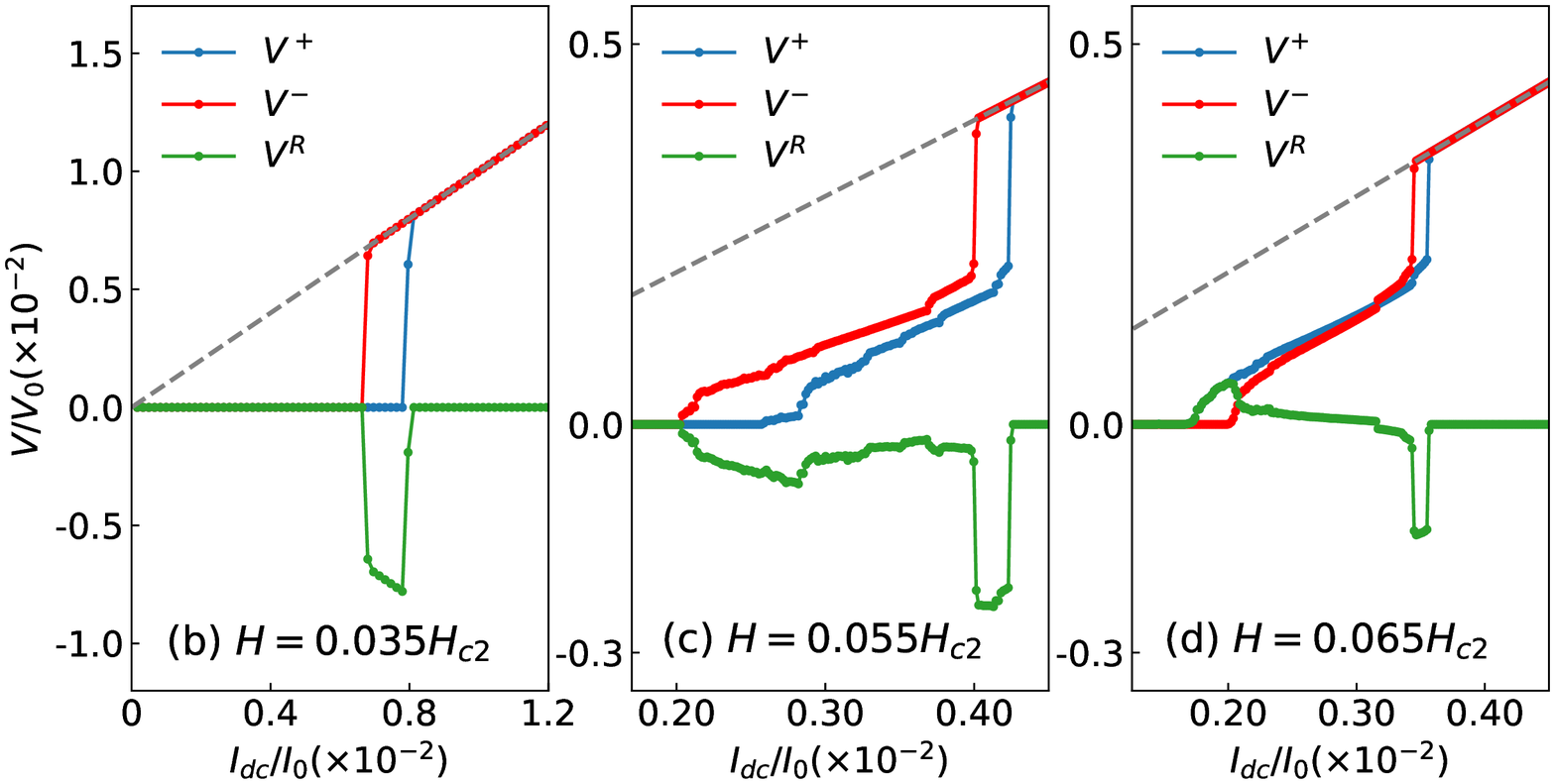}
\caption{(a) Phase diagram of ratchet voltage signals $V^{R}=V^{+}-V^{-}$ in
the current and field parametric range. The phase with negative signals (pink) is
present at high currents or low fields, and the phase with positive signals
(blue) appears at low currents and high fields. The thick red line
represents the transition line to the normal state driven by the counter-clockwise
currents. The lowest boundaries of the negative (left) and
positive (right) ratchet signals are corresponding to the critical current
lines for counter-clockwise and clockwise currents, respectively. Lower
panels show voltages and their differences as a function of external dc current of two
directions at magnetic fields $H=0.035H_{c2}$ (b), $0.055H_{c2}$ (c), and $%
0.065H_{c2}$ (d), where the dashed lines stand for the voltage in the
normal state. The voltage unit is $V_{0}=I_{0}/\protect\sigma $. As a criterion
for critical current, we take the onset of voltage above $10^{-5} V_{0}$.}
\label{fig_dc}
\end{figure}

In order to obtain the overall ratchet effect in a narrow
superconducting ring, we first systematically calculate the voltage signals
for dc currents of opposite directions. Figure \ref{fig_dc}(a) presents
the phase diagram of the ratchet signal as a function of the magnetic field $H$ and the
amplitude of the dc current $I_{dc}$. The amplitudes of the voltage induced by a
clockwise current is denoted as $V^{+}$ and that by a counter-clockwise current is denoted $V^{-}$.
The difference of the voltage values obtained for two opposite directions of the current $V^{R}=V^{+}-V^{-}$
is used to quantitatively describe the ratchet signal. Negative signals
(pink) are found at low fields ($H<0.060H_{c2}$) independent of the current value.
However, at high fields ($H>0.060H_{c2}$), positive signals (blue) emerge at low currents,
indicting a ratchet signal reversal. To the best of our knowledge, such a
ratchet reversal has not been reported to date in a pinning-free
superconducting system, neither experimentally nor in simulations.

To understand this phase diagram better, we show voltage-current
(V-I) characteristics at three magnetic fields $H=0.035H_{c2}$, $%
0.055H_{c2}$, and $0.065H_{c2}$ in the panels (b-d) of Fig. \ref{fig_dc}, respectively.
When the sample is driven to the normal state, the V-I characteristics follows ohmic behavior,
which is indicated by dashed lines in the lower
panels (b-d) of Fig. \ref{fig_dc}. Non-zero and zero voltage signals  below the
reference lines correspond to the dissipation state due to moving vortices
(flux-flow state) and the zero-resistance superconducting state,
respectively.

At the low field $H=0.035H_{c2}$, as shown in Fig. \ref{fig_dc} (b), in the dc
currents range $0.007<I_{dc}<0.008$, the system has zero resistance for the
clockwise current and turns to the normal state at counter-clockwise
currents. In this case, pronounced negative ratchet signals are generated
owing to the large voltage difference between the superconducting and the normal
phase. In other words, the superconducting and normal conducting states can be fully switched
by changing the direction of the applied current or magnetic field.
Interestingly, this is exactly the superconducting diode effect, similar to that
demonstrated in the artificial superlattice \cite{ando2020observation}, providing
a very economical way to fabricate a superconducting diode
from the perspective of material design.

At the intermediate field $H=0.055H_{c2}$, the ratchet signals are also all
negative [see Fig. \ref{fig_dc}(c)]. Panels (b) and (c) of Fig. \ref{fig_dc} indicate
that the sample with counter-clockwise current $-I$ (red) enters both the flux-flow
state and the normal state earlier than in the case of clockwise current $+I$ (blue),
yielding a negative ratchet signal. On the other hand, at a higher magnetic
field $H=0.065H_{c2}$ [see Fig. \ref{fig_dc}(c)], the sample in counter-clockwise
current $-I$ enters the flux-flow state later than in $+I$, which results in the
positive ratchet signal at small current, while the transition to the normal state
for $-I$ still occurs earlier, so the negative ratchet signal is also present at high
currents. At a high $-I$ the sample can transit to the normal phase while it can still stay
in the flux-flow \ state at the same value of $+I$. In this case,
the ratchet signal should be extremely large because of the high normal
state voltage, leading to pronounced negative signal at high currents in
the phase diagram in Fig. \ref{fig_dc}(a). We note that the total magnetic
field is the superposition of the external magnetic field and the field $H_{I}$ induced
by the current. For a current-carrying ring, the total field is
enhanced by the additional fields $H_{I}$ induced by $-I$ (counter-clockwise)
but suppressed by the fields of $+I$ (clockwise). Therefore, the superconducting ring
driven by $-I$ reaches the normal phase earlier than when driven by $+I$.

From Fig. \ref{fig_dc}(d) one sees that the positive ratchet signal
originates from the larger negative critical current at magnetic fields $%
H>0.060H_{c2}$. In equilibrium simulations with zero applied current, we
find that vortices can only penetrate the sample at fields $%
H>0.060H_{c2}$, which inspired us to study the underlining mechanism in
terms of vortex potential energy.

\subsection{Mechanism based on potential energy of a single vortex}

\begin{figure}[tbp]
\centering
\includegraphics[scale=0.42]{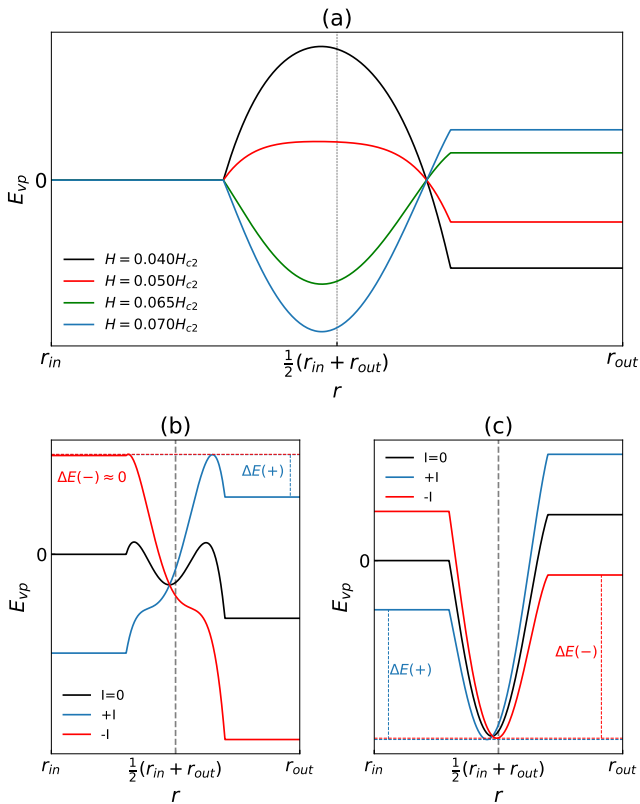} %
\caption{(a) Potential energy $E_{vp}$ of a single vortex inside the narrow
superconducting circular strip as a function of its radial position $r$ at
various magnetic field $H$ in the absence of external current. Lower panels
show the potential energy with the applied current $|I|=0.002$ at the magnetic
field $H=0.055 H_{c2}$ in (b) and $H=0.065 H_{c2}$ in (c), \textit{i.e.} in the
vicinity of the critical currents in Figs. \ref{fig_dc}(c) and (d), respectively.}
\label{fig_EvpH}
\end{figure}

The energy barrier for the penetration of vortices is known to determine the critical
current at which the superconductor enters the flux-flow state \cite%
{morgan2015measurement,pekker2011weber,maksimova1998mixed,maksimova1998mixed,likharev1971formation}%
. At equilibrium state, a higher energy barrier near the edge indicates that
a larger current is needed to drive vortices to the flux-flow state. In what
follows, we discuss critical current behavior in terms of the single vortex
potential in order to reveal the mechanism of the ratchet signal reversal.

The potential energy of a single vortex inside a symmetric superconducting
strip has been well studied and thoroughly explored \cite%
{maksimova1998mixed,likharev1971formation,stan2004critical,sanchez2007critical}%
. In general, vortex potential energy consists of four parts: (i) vortex
core energy, \ (ii) interaction energy between a vortex and its image, (iii)
interaction energy with magnetic field, and (iv) interaction energy with the
applied current. Following the ideas in these works, we perform conformal
transformation to the potential energy of a strip, and obtain the vortex
potential $E_{vp}$ of a narrow superconducting ring as a function of the
radial position $r$ of the ring:
\begin{equation}
\begin{split}
E_{vp}(r) = \frac{2\pi \rho}{r^{\prime}}ln\frac{sin(\pi y/w^{\prime })}{%
sin(\pi \xi ^{\prime }/w^{\prime})} + \frac{2\pi \rho H}{r^{\prime}}[(y-\frac{w^{\prime }}{2}%
)^{2} - \\ (\xi ^{\prime }-\frac{w^{\prime }}{2})]^{2} %
+ \frac{\Phi_0 J}{w^{\prime } r_{in}^{\prime}} (y^{\prime } - \frac{w^{\prime }}{2}),  \label{eq_Evp}
\end{split}
\end{equation}%
where $\rho $ is the superfluid stiffness, $\Phi_0$ is the flux quantum, and $J$ is the bias
current density. The scaling is changed due to the conformal transformation and the length is measured in units of
$\xi ^{\prime}=\ln \frac{\xi + r_{in}}{r_{in}}$: $y=\ln (r/r_{in}) / \xi ^{\prime}$,
$w^{\prime}=\ln(r_{out}/r_{in}) / \xi ^{\prime}$, and $r^{\prime}=r/ \xi ^{\prime}$. The logarithmic
divergence near the edge is cut off by vortex core length scale $%
\xi = \xi(T=T_{0})$ \cite{maksimova1998mixed,likharev1971formation}, which leads to the flat platform
near sample edges. As vortices are unstable in the vicinity of the boundary and would exit from the edge
or enter the sample quickly, this approach remains valid for the purpose of our discussion. Without
the loss of generality, we have absorbed the vortex core energy into the cut-off length
scale. In Fig. \ref{fig_EvpH}(a), we plot the vortex potential at a few selected magnetic
fields without external currents for a superconducting ring. Compared to that of straight
nanowires \cite{morgan2015measurement,pekker2011weber}, the vortex potential of a
ring-shaped superconductor inherits asymmetric energy barriers near its inner and outer edges.
The vortex potential described here is qualitatively consistent with the analytical
findings in the annular ring in a uniform applied field using the London approach \cite%
{kogan2004properties,kirtley2003fluxoid}. In low-current regime, the total potential is
only tilted due to the Lorentz force and the asymmetric energy barriers at two
edges are preserved, as shown in the lower panels of Fig. \ref{fig_EvpH}.

At low magnetic fields where $H<0.060H_{c2}$, the energy barrier indicates
that in equilibrium vortices cannot enter the sample,
which is consistent with our numerical observations. Above the critical current,
current-induced vortices driven by Lorentz force could enter at one edge and then
exit from the other edge. In the present setup, vortices should move from the inner
edge towards the outer edge at a counter-clockwise current ($-I$), and vice versa.
In Fig. \ref{fig_EvpH}(b), the energy barrier for vortex entry with $+I$ ($-I$) is indicated by $\Delta E(\pm)$
by the blue (red) lines. Since  $\Delta E(-) < \Delta E(+)$, the  flux-flow
state is easier to reach for $-I$, leading to a lower critical current for $-I$, and accordingly a
negative ratchet signal.

However at higher fields ($H > 0.060H_{c2}$), the energy barriers are replaced
by potential wells. Now vortices can appear inside the sample even in the
absence of external current. In Fig. \ref{fig_EvpH}(c), we again  use
$\Delta E(\pm)$ to denote the well depth associated to the current $\pm I$. In contrast to
the low-field case, we now have $\Delta E(-) > \Delta E(+)$. Therefore, vortices
can exit from the inner edge easier than from the outer one, yielding a higher
critical current for $+I$. In other words, at these fields, the ratchet signal turns to be
positive. With further increase of the external current, approaching
the current above which the sample transits to the normal state, the energy well
difference at two edges becomes far less important due to the dominance
of the Lorentz force. Therefore, in the high-current regime, the effect of the current-induced
field discussed in Sec. III(A) dominates and the negative ratchet signal reappears.

\section{ Ratchet effects generated by ac currents}

\begin{figure}[tbp]
\centering
\includegraphics[scale=0.5]{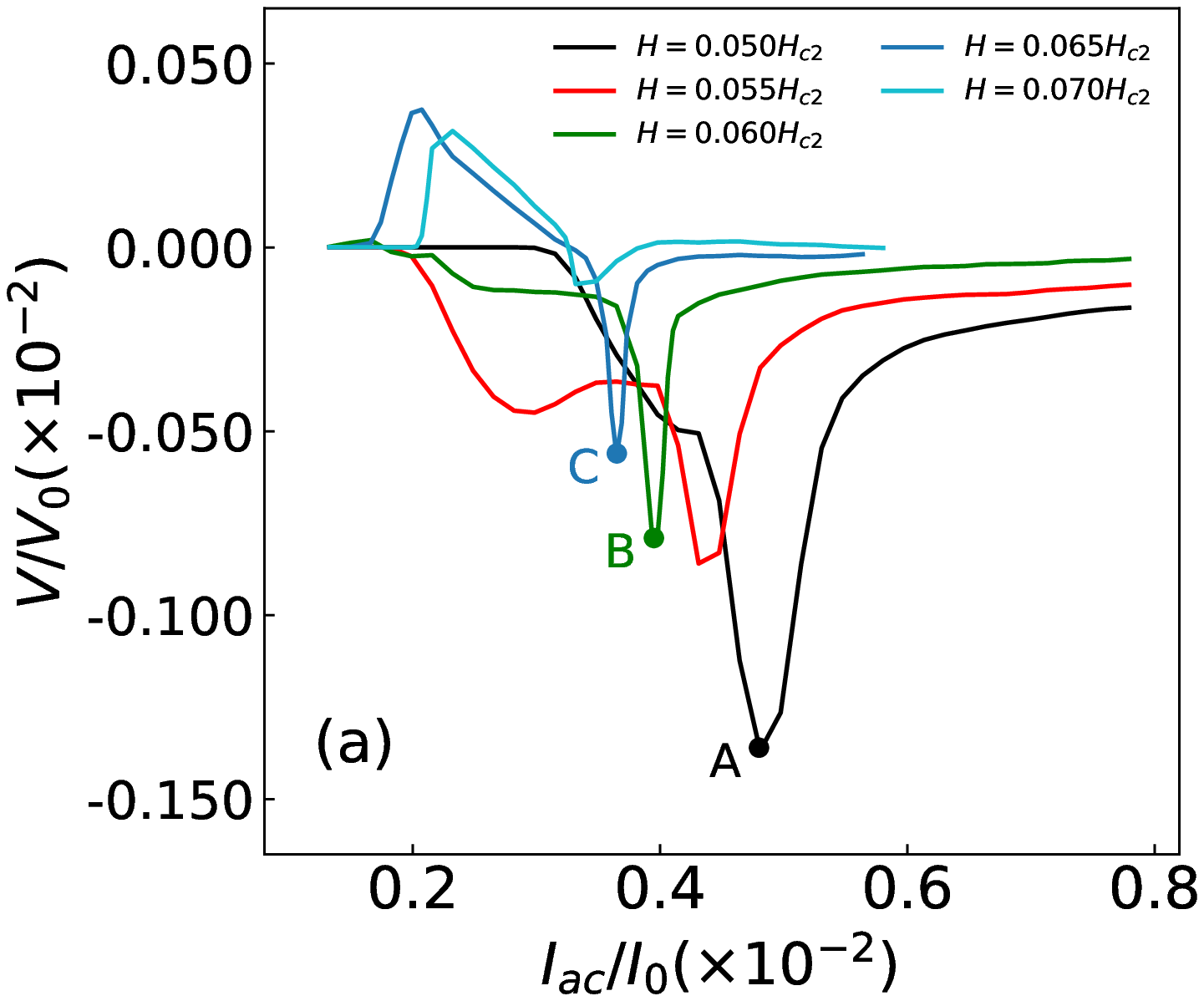}
\includegraphics[scale=0.5]{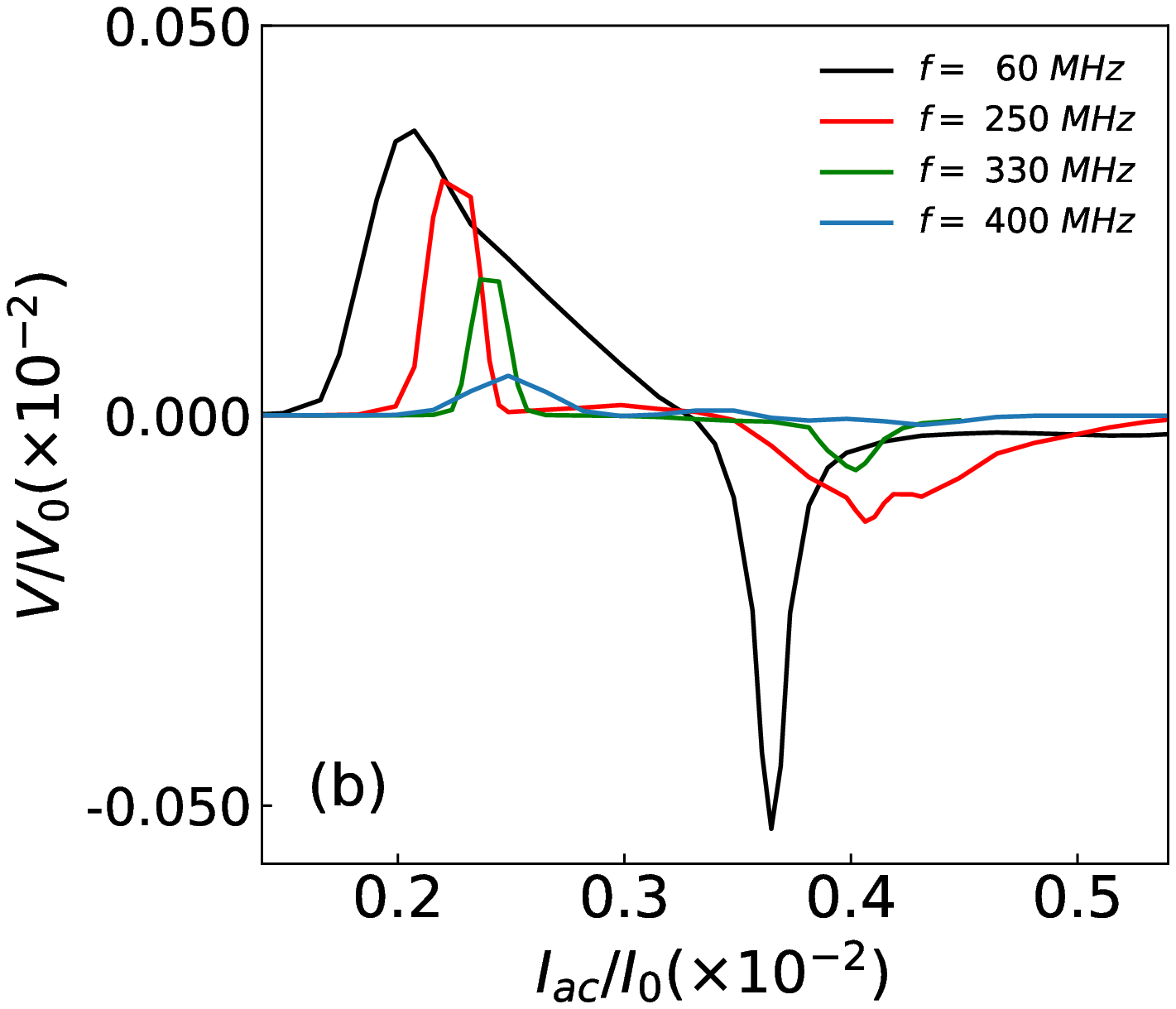}
\caption{Mean dc voltage as a function of the amplitude of the applied
ac current at various magnetic fields (a), and at various frequencies (b).
$I_{ac}$ is the amplitude of the applied ac current with a fixed frequency
$f=60$ MHz. A, B and C mark three prominent negative signal peaks at $H=0.050$, 0.060
and 0.065$H_{c2}$, respectively. The magnetic field for (b) is kept fixed at $H=0.065H_{c2}$.}
\label{fig_VI_H}
\end{figure}

Next we study the response of the ring to the ac currents with zero mean.
We apply a sinusoidal current $I(t)=I_{ac}\sin \left( \frac{2\pi }{P}t\right) $
to the sample, where $P=8000t_{GL}$. According to $t_{GL}$ in typical
MoGe samples \cite{wang2017parallel}, the related frequency is about $%
f=60.0$ MHz. Our simulations suggest that such a frequency
allows vortices to move across the strip within $P/2$ for a wide range of
amplitudes $I_{ac}$. The mean dc voltage is obtained by averaging over five
periods $P$. As shown in Fig. \ref{fig_VI_H}(a), the rectified dc voltage is
clearly observed in a finite range of magnetic fields. At low fields, only
negative ratchet signals are observed. The ratchet signal decreases
monotonously when the field is increased. With increasing magnetic field, we
find that a positive ratchet signal arises at low currents while the
negative signal remains in the high current range.

\begin{figure}[tbp]
\centering
\includegraphics[scale=0.5]{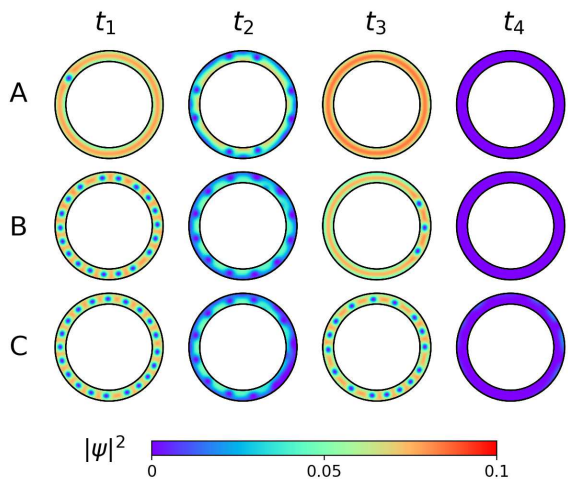}
\caption{Snapshots for the evolution of the Cooper-pair density $|\protect\psi |^{2}$.
Images in rows A, B and C present respectively $|\protect\psi |^{2}$ at peak points
A, B and C in Fig. \ref{fig_VI_H}(a) at $t_{1}=0$, $t_{2}=0.3P$, $t_{3}=0.5P$,
and $t_{4}=0.8P$ within one period.}
\label{fig_evo}
\end{figure}

To understand the details of the negative ratchet process at high current, we
plot in Fig. \ref{fig_evo} snapshots of the Cooper-pair density at different times $%
t=0,0.3,0.5,$ and $0.8P$ within one period, at the three ratchet signal peaks
A, B, and C marked in Fig. \ref{fig_VI_H}(a). After closer inspection, we find that
the current $I_{pk}$ at which the ratchet peak signal occurs is close to the dc current
for the transition to the normal state. It implies that if the ac amplitude is comparable to the
dc critical current to the normal state, the sample can transit to normal phase within one
period. As shown in the rightmost panel of Fig. \ref{fig_evo} at $t_{4}=0.8P$%
, the sample indeed enters the normal state with vanished order parameter
$\left\vert \Psi \right\vert ^{2}=0$. At the other three times, all earlier
than $t_{4}$, vortices are clearly present, corresponding to the flux-flow
state. For detailed visualization of the vortex dynamics, the corresponding videos, Video-S1(A/B/C),
are provided in supplemental material \cite{Supplemental}. One can see from both
the snapshots and the videos that during the first half-period, individual vortices
can be clearly seen, suggesting that the sample remains in the flux-flow state.
However, in the second half-period, the vortex cores overlap very quickly
and the sample transits to the normal state before the end of the period. Due to the
significant voltage difference between the two states, maximum ratchet effect occurs.
Below $I_{pk}$ the sample stays less time in normal state in the second half-period hence the
ratchet signal decreases. Above $I_{pk}$, the normal state can be accessed
in both half-periods, consequently the ratchet effect also becomes smaller.
That is, the large difference of the voltages in the time intervals in which the
sample stays in the normal phase and flux-flow state within one period results
in the peak structure of the ratchet signal. In principle, current above which
the sample is in the normal state decreases with the magnetic field, so the peak shifts to
the low ac amplitude range. Accordingly, the corresponding negative signals
should also become smaller, in agreement with our numerical observations in
Fig. \ref{fig_VI_H}(a).

One may notice that in Fig. \ref{fig_evo} vortex density is different
at $t_{1}=0$ and $t_{3}=0.5P$ when the external currents are both temporarily
zero. Generally speaking, we have a larger vortex density at $t_{1}=0$.
Before $t_{1}$, the sample just experiences the transition from the normal
phase to the flux-flow state. Due to the non-adiabatic process, even at
transient zero current, the vortex does not have sufficient time to be
expelled from the sample, which leads to a higher vortex density.

The mechanism of the positive signals that emerge at lower ac amplitude is
the same as that seen in the case of dc currents in Sec. III. As shown in 
Video-S2 in the supplemental material \cite{Supplemental}, for $I_{ac}=0.002I_{0}$
at $H=0.065H_{c2}$, in the first half-period ($+I$), the vortices can exit
at the inner edge, while in the second half-period ($-I$), the vortices
are locked inside the sample despite a finite Lorentz force. That is,
in the first half-period, the sample enters the flux-flow state, while in
the second half-period, the sample is still in the zero-resistance state,
which yields the positive ratchet signals.

The effect of ac frequency on the ratchet signal is also studied.
Without the loss of generality, the magnetic field was set to $H=0.065H_{c2}$.
Fig. \ref{fig_VI_H}(b) shows the mean voltage at various frequencies. The
ratchet signals are more pronounced at low frequencies. With increasing
frequency, both negative and positive signals decrease, and they gradually
disappear at high frequencies. A video for vortex behavior at
the same field and ac amplitude but at a higher ac
frequency $f=400$ MHz is provided in Video-S3 in the supplemental
material \cite{Supplemental}. At a high frequency, the ac current period is
so short that the vortices could not cross the ring during both the first
and second half-period, and can only oscillate inside the superconductor.
In this case, the vortex oscillation generates subtle difference in the
voltages for two opposite current directions, hence the ratchet effect is
strongly suppressed.

\begin{figure}[tbp]
\centering
\includegraphics[scale=0.5]{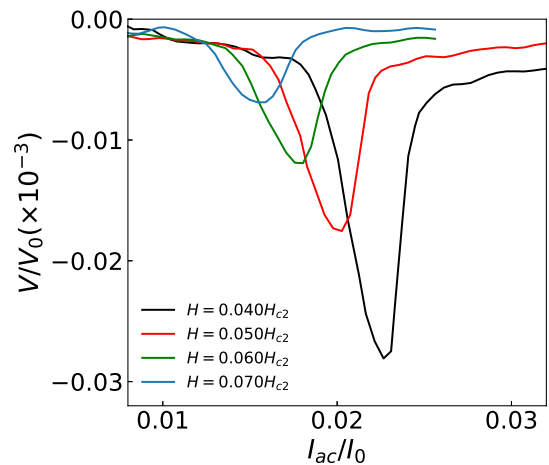}
\caption{Mean dc voltage as a function of the amplitude of applied ac
current at various magnetic fields, for a superconducting ring with
larger width. The frequency of the applied current was kept fixed at $f=60$ MHz.
}
\label{fig_VI_wd}
\end{figure}

We have also simulated the response to the ac current in a wider ring-shaped
superconductor.  Figure \ref{fig_VI_wd} shows the ratchet voltage signal for
a similar superconducting ring of size $r_{out}=60\xi(0) $ and $%
r_{in}=30\xi(0) $. The frequency of the applied ac current was kept at $f=60$ MHz.
With the increase of the ring width $w$, the reversal of the ratchet signal would vanish,
consistent with the experimental observation in the S20 sample of \cite{ji2017ratchet}. The
reason for the vanishing positive ratchet signal in a wider superconducting ring is that
for large enough width, more than one vortex can be presented along the
radial direction of the sample. Vortex-vortex interaction would not be
negligible and the above picture for the single vortex potential in Sec. III
B breaks down. Due to their repulsive interaction, vortices prefer to stay
away from others, thus it is more difficult for the vortices to move to
the inner edge (high vortex density) than towards the outer edge (dilute vortex
density), which eventually results in the negative signal in all cases, excluding
the possibility of the positive ratchet signal.

The ratchet effect in the pinning-free system with an asymmetric edge barrier
here is stronger than those found in superconductors with asymmetric pinning
potentials \cite{reichhardt2015reversible,adami2015onset}. In the latter
case, the ratchet effects are generated by the difference in the velocities
of vortices driven by opposite currents, in comparison with the former where
the rectified signal is generated by the voltage difference between
the flux-flow state and the zero-resistance state. As a consequence, the rectified
voltage signal due to asymmetric pinning can be one or two orders of magnitude smaller
than in case of asymmetric sample edge \cite%
{adami2015onset,wu2015special,ji2016vortex,ji2017ratchet}.

It has been reported in the literature that a rough edge can affects vortex
dynamics  alone \cite{cerbu2013vortex}.  In this paper, the sample boundary is assumed to be
perfect, \textit{i.e.} without local defects, so that we can focus exclusively on the effect of the asymmetric
geometry on the ratchet signal. To confirm the present observation in experiments, we would suggest that both the clean superconducting ring with minimal edge roughness and
the leads with very low dissipation should be fabricated to avoid unnecessary
noise signals. On the other hand, a real sample always contains intrinsic, weak random  pinning centers. However, the effects of this kind of pinning centers for the two opposite vortex motion currents are nearly the same, so the main picture proposed in this paper would not be affected.

Finally, as the signal reversal decreases with increasing sample width,
it is also crucial that the ring width should be comparable to the vortex size in a finite range
of temperature. With the  ring width $w=12 \xi(0)$ used in this paper, the reversible ratchet effect phenomenon can be convincingly observed
in the temperature range of $(0.75, 0.97)T_c$.

\section{Conclusion}

In summary, we systematically studied the ratchet effect in a narrow
superconducting ring using TDGL equations. The dc voltages in the presence
of external dc currents at various magnetic fields were calculated first. We
found a reversible ratchet effect using dc currents of opposite directions.
The superconducting diode effect is also observed. Strong
negative ratchet signals are found in the high current regime in a wide
range of magnetic fields. Surprisingly, at high fields, a positive ratchet
signal appears in the low current regime. It is shown in numerical
simulations that different critical currents for two polarities of the current are the
origin of the observed positive ratchet signal. It is further revealed that
this unusual phenomenon can be attributed to the asymmetric vortex potential
due to the ring-shaped structure.

Rectified voltage signals are also found
when applying ac currents to the system. We observed pronounced negative
voltage signals in a broad range of external fields while the positive
ratchet signals were also observed in the weak ac amplitude regime. Further
investigations suggest that positive ratchet signals observed with ac
and dc current share the same origin. It is also demonstrated that with
increasing ac frequency and ring width, the unusual positive ratchet signals
weaken and eventually vanish.

The pronounced ratchet signals observed in the pinning-free but geometrically
asymmetric system are generated from the switch of various phases, namely the zero-resistance phase, dissipative
phase (flux-flow state), and the normal phase when
the polarity of applied current is changed. Therefore, these signals are by
default larger than those caused by the different vortex velocities within the
same phase in systems with asymmetric pinning potentials. As a result, the large
and reversible ratchet signal seen in our simulations should stimulate further
experimental investigations and its use in superconducting circuits or devices, including
superconducting diodes and single-photon detectors.

\textbf{ACKNOWLEDGEMENTS} We are grateful to G. Berdiyorov for useful suggestions and
comments. Q.H.C. thanks Beiyi Zhu for helpful discussions during the early stage of this work.
This work is supported in part by the National Key Research and Development
Program of China (No.2017YFA0303002) (Q.H.C. and J.J.) and (2018YFA0209002) (Y.L.W.);
and the National Natural Science Foundation of China (Nos. 11834005, 11674285, 61771235,
and 61727805). Z.L.X. acknowledges support by the U. S. Department of Energy, Office of
Science, Basic Energy Sciences, Materials Sciences and Engineering and the National Science
Foundation under Grant No. DMR-1901843. F. M. Peeters and M. V. Milo\u{s}evi\'{c}
acknowledge support by the Research Foundation - Flanders (FWO).

\end{document}